\documentclass[onecolumn,showkeys,preprintnumbers,aps,a4paper,amssymb,prd,superscriptaddress,nofootinbib]{revtex4-2}
\usepackage{comment}
\usepackage{graphicx}
\usepackage{epsf}
\usepackage{bm}
\usepackage{amsmath}
\usepackage{amsfonts}
\usepackage{amssymb}
\usepackage{epstopdf}
\usepackage{color}
\usepackage[dvipsnames]{xcolor}
\usepackage{verbatim}
\usepackage{multirow}
\usepackage{soul}
\usepackage{physics}
\usepackage{bm}

\usepackage[width=0.00cm, height=0.00cm, left=1.50cm, right=1.50cm, top=2.00cm, bottom=2.00cm]{geometry}
\usepackage{microtype}
\usepackage[colorlinks = true,
            linkcolor = teal,
            urlcolor  = black,
            citecolor = teal,
            anchorcolor = blue]{hyperref}
\usepackage[capitalize]{cleveref}
\usepackage[normalem]{ulem}
\usepackage{enumitem}
\usepackage{booktabs}

\usepackage{lipsum}

\makeatletter\let\expandableinput\@@input\makeatother

\begin{document}

\begin{center}
		\vspace{0.4cm} {\large{\bf  Constraints on Spatial Curvature and Dark Energy Dynamics in the $w$CDM Model from DESI DR1 and DR2
}} \\
		\vspace{0.4cm}
		\normalsize{ Manish Yadav $^1$, Archana Dixit$^2$, M. S. Barak$^3$,  
         Anirudh Pradhan$^4$ }\\
		\vspace{5mm}
		\normalsize{$^{1,3}$ Department of Mathematics, Indira Gandhi University, Meerpur, Haryana 122502, India}\\
		\normalsize{$^{2}$ Department of Mathematics, Gurugram University, Gurugram- 122003, Harayana, India.}\\ 
        \normalsize{$^{4}$ Centre for Cosmology, Astrophysics and Space Science (CCASS), GLA University, Mathura-281 406, Uttar Pradesh, India}\\
		\vspace{2mm}
		$^1$Email address: manish.math.rs@igu.ac.in\\
		$^2$Email address: archana.ibs.maths@gmail.com\\
            $^3$Email address: ms$_{-}$barak@igu.ac.in\\
            $^4$Email address: pradhan.anirudh@gmail.com\\
\end{center}

{\noindent {\bf Abstract.}
 
In this study, we investigate the $w$CDM dynamical dark energy model with spatial curvature utilizing the recently released DESI Collaboration data (DR1 and DR2) in conjunction with other observational probes such as BBN, Observational Hubble Data (OHD), and Pantheon Plus (PP).  Our investigation attempts to discover which DESI dataset gives a better match to the $w$CDM framework and assess the impact of spatial curvature on cosmological constraints. We find that the cosmic curvature parameter, $\Omega_k$, disfavors the cosmological constant for the DR2+BBN and DR2+BBN+OHD data combinations. However, the deviation from the cosmological constant remains below the $1\sigma$ level, indicating a mild preference for a open universe. In contrast, when using the DR1 based combinations namely DR1+BBN and DR1+BBN+OHD—the deviation from the cosmological constant increases to approximately $1.2\sigma$, suggesting a slightly stronger indication of a open geometry. Also, the best-fit values of the Hubble constant ($H_0$) obtained from the DR1+BBN+OHD+PP and DR2+BBN+OHD+PP combinations within the dynamical dark energy model are consistent with the results reported by the Planck Collaboration. Our findings provide constraints on the dark energy EoS parameter $ w_{\mathrm{}0}$, reveal a mild but notable deviation from the vacuum energy ($w=-1$) scenario at a significance level $1.8\sigma$ from DR2+BBN+OHD+PP and  $0.5\sigma$ from DR1+BBN+OHD+PP, both favoring the quintessence region of dark energy. Furthermore, the key physical distance measures $D_H$, $D_V$, and $D_M$ show better consistency with our model when analyzed with the DR2 data.

\section{Introduction}

In the past decades, the Lambda Cold Dark Matter ($\Lambda$CDM) model has served as the standard framework in cosmology, providing remarkably consistent with observational cosmological and astrophysical data across multiple scales and cosmic epochs \cite{ref1,ref2,ref3,ref4,ref5,ref5a}. However, the last decade has seen emerging discrepancies in certain important cosmological parameters, such as the Hubble parameter and the density parameter of curvature. These discrepancies have spurred the scientific community to pursue alternative explanations, whether by invoking new physical theories or by examining potential sources of systematic errors in the observations, see ref.\cite{ref6, ref7,ref8,ref8a,ref8b,ref8c}. Several studies have been made to investigate novel physics as a potential solution to cosmological tensions. Most of these adopt a bottom-up strategy, assuming the existence of an effective field theory that can explain why the standard cosmological model is constrained by diverse and robust datasets. In this system, addressing model degeneracy is dependent on improving the fit to the data. 

A fundamental geometric property of the cosmos is its spatial curvature, indicated by $k$.  Given that most widely accepted models of early Universe inflation predict spatial flatness ($k=0$), any discovery of nonzero spatial curvature-hereafter simply referred to as curvature-would have substantial consequences for our understanding of the early Universe and its subsequent history.  Several approaches have been proposed for measuring and constraining cosmic curvature. A model-independent technique, developed by \cite{ref9}, predicts the curvature with an accuracy of $6 \times 10^{-3}$ utilizing strong lensing time delays and supernova distance measurements.  Ref. \cite{ref10} emphasizes the importance of curvature in cosmological testing of General Relativity (GR), stating that its inclusion eliminates spurious deviations from GR and modifies the permitted parameter space for growth. According to ref. \cite{ref11}, identifying curvature with high accuracy is troublesome since it may not be achievable without assuming precise forms for dark energy (DE) evolution and using conventional $\Lambda$CDM parameter values. \\

 The author in Ref. \cite{ref12} proposes a test-particle method for measuring curvature, which requires at least six particles in a general setting and at least four particles in vacuum. The use of gravitational-wave standard sirens and Cosmic Chronometers (CC)  in model-independent determinations of cosmic curvature has been proposed in several numerical studies. Ref. \cite{ref13} and \cite{ref14} illustrate their promise, with \cite{ref14} suggesting that the DECIGO space-based observatory could achieve trustworthy limits on curvature. Additional applications of gravitational waves and CC  are examined by ref. \cite{ref15} and \cite{ref16}, where \cite{ref16} suggests that combining gravitational waves with strong gravitational-lensing systems enables measurement of cosmic curvature. Building on this work, ref. \cite{ref17} and \cite{ref18} applied model-independent approaches to the same observational datasets—cosmic chronometers (CC) and gravitational-wave standard sirens—using machine learning, and found that these methods show considerable potential for accurately measuring cosmic curvature. The Planck 2018 analyses of CMB temperature and polarization spectra, with the Plik likelihood,  point toward a closed universe at $>3\sigma$ significance,\cite{ref3,ref19,ref20,ref21}, reaching $4\sigma$ with physical curvature density considerations \cite{ref22}. This arises primarily from a lensing excess in the temperature data ($A_{\rm lens}$ problem \cite{ref23,ref24}). Complementary analyses using CamSpec \cite{ref25,ref26} also indicate a closed geometry at $>99\%$ confidence, a result that persists in the Planck PR4 temperature-only dataset \cite{ref27}. Several other studies addressing cosmic curvature are also presented in refs. \cite{ref28,ref29,ref30,ref31,ref32,ref33,ref34,ref35,ref36,ref37}.\\

In one of the most recent works, Bikash R. Dinda \cite{ref38} investigated a dynamical dark energy scenario based on the Chevallier--Polarski--Linder (CPL) parametrization using the combined dataset of DESI DR2, CMB, and Pantheon+ (PP) observations, and reported indications of a non-zero cosmic curvature. In a subsequent analysis, Dinda \cite{ref39} further examined the CPL model with an extended data combination including CC+DR2+PP+ the sound horizon at the drag epoch ($r_d$) + growth rate measurements ($f\sigma_8$). This analysis yielded a non-zero mean value of $10^2\Omega_k = 1.8$ for the CPL model and $10^2\Omega_k = 0.8$ for the $w$CDM model with the same dataset. In parallel, Pengu \cite{ref40} explored a range of dynamical dark energy models---namely $w$CDM, exponential, CPL, JBP, and interacting dark energy (IDE)---using DESI DR2, Supernova (SN), and Time-Delay (TD) lensing data, and similarly found evidence for a small but non-zero curvature of $\Omega_k \approx 0.1$. More recently, further extensions of these analyses have incorporated CC data into the observational combinations, and they found consistent evidence for non-zero cosmic curvature, reinforcing the case for exploring dynamical dark energy models \cite{ref41}. In contrast, Shi-Fan Chen \cite{ref42} examined the $\Lambda$CDM+$\Omega_k$+$M_\nu$ framework and discovered a minor negative spatial curvature.  Motivated by recent studies indicating a possible deviation from spatial flatness within dynamical dark energy scenarios, we extend the investigation by including the $w$CDM+$\Omega_k$ model. For this goal, we employ high-precision data from the DESI project, notably the first (DR1) and second (DR2) data releases, in conjunction with complementing cosmological probes.  A special emphasis is placed on determining which dataset, DR1 or DR2, provides more balanced and robust restrictions on cosmic curvature and the dynamical character of dark energy.\\
In this study, we introduce the dynamical dark energy model (wCDM) with cosmic spatial curvature and investigate its consequences for the Universe's evolution.  Our primary focus in this study is the presence of curvature effects on the other cosmological parameters in the DR1 and DR2 datasets. We also examine their combination with other complementary observational probes, such as CC and Pantheon+ supernova measurements, to determine which dataset or combination provides the most robust and balanced constraints on the $w$CDM model, shedding light on the interaction of spatial curvature and dark energy dynamics.  This complete approach allows for a more precise assessment of the function of curvature and compares the relative strengths of various data releases in constraining cosmological theories. \\

The basic framework of this paper is as follows. In Section I, we provide a review of the literature related to the $w$CDM model with spatial curvature. Section II outlines the observational datasets employed in this work, including OHD, Pantheon+ (PP), DESI DR1, and DR2, along with the methodology adopted for the analysis. In Section III, we present our results and discuss the implications of the findings. Finally, Section IV summarizes the main conclusions of the study.
\section{MODEL, DATA AND METHODOLOGY}\label{sec2}


\label{sec:datasets}

In continuation of our previous work \cite{ref43}, we consider a dynamical dark energy model with spatial curvature, denoted as the $w$CDM + $\Omega_k$ model. The governing Friedmann equation for this model is given by

\begin{equation}
    \frac{H^2(z)}{H_0^2}  =\Omega_{\text{r}0}[1+z]^{4}+\Omega_{\text{m}0}[1+z]^{3}+ \Omega_{\text{k}0}[1+z]^{2} + \Omega_{\rm de0}[1+z]^{3(1+w_{\rm de0})},
\end{equation}

Where $H_0$ is the present value of the Hubble parameter, and $\Omega _{k0}$
denote the present-day curvature density parameter. The curvature density is related to the spatial curvature  $k$ of the universe by $\Omega _{k0} = \frac{-c k}{H_0^2}$, where c is the speed of light in vacuum. The geometry of the universe depends on the sign of cosmic curvature of the universe ($k$), which depends on an open universe ($k < 0$), a closed universe ($k> 0$), and a flat universe $(k = 0$).\\

In our analysis, we neglect the contribution of the radiation density parameter, since radiation has a negligible effect at late times. Consequently, Eq. (1) reduces to
\begin{equation}
    \frac{H^2(z)}{H_0^2}  =\Omega_{\text{m}0}[1+z]^{3}+ \Omega_{\text{k}0}[1+z]^{2} + \Omega_{\rm de0}[1+z]^{3(1+w_{\rm de0})},
\end{equation}

Here, $\Omega_{m0} $ and $\Omega_{de0}$ denote the present-day density parameters of matter and dark energy, respectively. The present-day values of these parameters satisfies the energy budget restriction condition in the equation 
   $ \Omega_{m0}+ \Omega_{k0} + \Omega_{de0} = 1$.\\

   The datasets and methodology used are as follows:

 \textbf{Dark Energy Spectroscopic Instrument}: The first-year data release by the DESI Collaboration (DRI) \cite{ref44,ref45,ref45a,ref45b}, which includes seven tracers with redshift bands: BGS ($0<z<0.4$), LRG1($0.4<z<0.6$), LRG2($0.6<z<0.8$), LRG+ELG($0.8<z<1.1$), ELG($1.1<z<1.6$), QSO ($0.8<z<2.1$), Ly$\alpha$ QSO ($1.77<z<4.16$). These redshift bands, along with the corresponding comoving distances, are reported in Table 1 of ref.\cite{ref46}. Subsequently, the DESI Collaboration released the second data release (DR2), which extended the analysis to nine tracers, as reported in ref.\cite{ref47}. These tracers provide more precise measurements of three important quantities,  $D_H(z)/r_{\rm d}$, $D_{M}(z)/r_{\rm d}$, and  $D_{v}(z)/r_{\rm d}$ with effective redshift are defined as:
\begin{itemize}
\item $\textbf{Transverse Comoving Distance with Cosmic Curvature of the Universe:}$\\

$D_{M}(z)=  \int_0^z \text{d}z' {c \over H(z')}S_{\kappa}(x)$. Where $c$ is the speed of light and \\

$S_{\kappa}(x) =
\begin{cases}
\dfrac{\sin\!\left(\sqrt{-\Omega_{\kappa}}\,x\right)}{\sqrt{-\Omega_{\kappa}}}, & \Omega_{\kappa} < 0, \\[10pt]
x, & \Omega_{\kappa} = 0, \\[10pt]
\dfrac{\sinh\!\left(\sqrt{\Omega_{\kappa}}\,x\right)}{\sqrt{\Omega_{\kappa}}}, & \Omega_{\kappa} > 0.
\end{cases}$

\item\textbf{Hubble Distance}:
$ D_H(z) = \frac{c}{H(z)}$,

\item\textbf{Angle-Averaged  Distance}:
$D_V(z) \equiv \left[z D^2_M(z) D_H(z)\right]^{1/3}$
 
\end{itemize}
where $r_{\rm d}=\int_{z_{\rm d}}^\infty \frac{c_{\rm s}\text{d}z}{H(z)}$  and  ($c_{\rm s}$) are sound horizon at the drag redshift ($z_{\rm d}$) and sound speed  of the baryon-fluid of the universe, respectively .\\

We define $\chi^2$ function on the DR1 and DR2 measurements with covaraince metrix:

\begin{equation}
	\chi^2_{\text{DR1}} =  \sum_{i}^{} \frac{[\left( \frac{D_V}{r_d} \right)^{\text{obs}}(z_i)-\left( \frac{D_V}{r_d} \right)^{\text{th}}(z_i)]^2}{\sigma^2_{\left( \frac{D_V}{r_d} \right)^{\text{obs}}(z_i)}}+ \Delta D_i \left(C_{\text{DR1}}^{-1}\right) \Delta {D_i}^T.
\end{equation}

\begin{equation}
	\chi^2_{\text{DR2}} =  \sum_{i}^{} \frac{[\left( \frac{D_V}{r_d} \right)^{\text{obs}}(z_i)-\left( \frac{D_V}{r_d} \right)^{\text{th}}(z_i)]^2}{\sigma^2_{\left( \frac{D_V}{r_d} \right)^{\text{obs}}(z_i)}}+ \Delta D_i \left(C_{\text{DR2}}^{-1}\right) \Delta {D_i}^T.
\end{equation}

where,  $C_{\text{DR1}}^{-1}$ \cite{ref48} and $C_{\text{DR2}}^{-1}$ are the inverse of the covariance matrix 

\[
C^{-1}_{\text{DR1}} =
\begin{pmatrix}
19.95 & 3.64  & 0     & 0     & 0     & 0     & 0     & 0     & 0     & 0 \\
3.64  & 3.35  & 0     & 0     & 0     & 0     & 0     & 0     & 0     & 0 \\
0     & 0     & 11.86 & 2.66  & 0     & 0     & 0     & 0     & 0     & 0 \\
0     & 0     & 2.66  & 3.37  & 0     & 0     & 0     & 0     & 0     & 0 \\
0     & 0     & 0     & 0     & 15.03 & 4.68  & 0     & 0     & 0     & 0 \\
0     & 0     & 0     & 0     & 4.68  & 9.62  & 0     & 0     & 0     & 0 \\
0     & 0     & 0     & 0     & 0     & 0     & 2.62  & 1.91  & 0     & 0 \\
0     & 0     & 0     & 0     & 0     & 0     & 1.91  & 7.06  & 0     & 0 \\
0     & 0     & 0     & 0     & 0     & 0     & 0     & 0     & 1.47  & 3.86 \\
0     & 0     & 0     & 0     & 0     & 0     & 0     & 0     & 3.86  & 44.79\\
\end{pmatrix}
\]

\[
\mathbf{C}_{\rm DR2}^{-1} =
\left(
\begin{array}{rrrrrrrrrrrrrrrr}
45.43 & 8.19  & 0 & 0 & 0 & 0 & 0 & 0 & 0 & 0 & 0 & 0 & 0 & 0 & 0 & 0 \\
8.19  & 7.01   & 0 & 0 & 0 & 0 & 0 & 0 & 0 & 0 & 0 & 0 & 0 & 0 & 0 & 0 \\
0 & 0 & 38.15 & 8.26 & 0 & 0 & 0 & 0 & 0 & 0 & 0 & 0 & 0 & 0 & 0 & 0 \\
0 & 0 & 8.26  & 10.97 & 0 & 0 & 0 & 0 & 0 & 0 & 0 & 0 & 0 & 0 & 0 & 0 \\
0 & 0 & 0 & 0 & 37.79 & 11.87 & 0 & 0 & 0 & 0 & 0 & 0 & 0 & 0 & 0 & 0 \\
0 & 0 & 0 & 0 & 11.87 & 22.63 & 0 & 0 & 0 & 0 & 0 & 0 & 0 & 0 & 0 & 0 \\
0 & 0 & 0 & 0 & 0 & 0 & 52.34 & 17.14 & 0 & 0 & 0 & 0 & 0 & 0 & 0 & 0 \\
0 & 0 & 0 & 0 & 0 & 0 & 17.14 & 32.46 & 0 & 0 & 0 & 0 & 0 & 0 & 0 & 0 \\
0 & 0 & 0 & 0 & 0 & 0 & 0 & 0 & 11.32 & 5.90 & 0 & 0 & 0 & 0 & 0 & 0 \\
0 & 0 & 0 & 0 & 0 & 0 & 0 & 0 & 5.90 & 14.41 & 0 & 0 & 0 & 0 & 0 & 0 \\
0 & 0 & 0 & 0 & 0 & 0 & 0 & 0 & 0 & 0 & 12.18 & 7.60 & 0 & 0 & 0 & 0 \\
0 & 0 & 0 & 0 & 0 & 0 & 0 & 0 & 0 & 0 & 7.60 & 25.22 & 0 & 0 & 0 & 0 \\
0 & 0 & 0 & 0 & 0 & 0 & 0 & 0 & 0 & 0 & 0 & 0 & 2.30 & 1.57 & 0 & 0 \\
0 & 0 & 0 & 0 & 0 & 0 & 0 & 0 & 0 & 0 & 0 & 0 & 1.57 & 4.31 & 0 & 0 \\
0 & 0 & 0 & 0 & 0 & 0 & 0 & 0 & 0 & 0 & 0 & 0 & 0 & 0 & 4.27 & 9.28 \\
0 & 0 & 0 & 0 & 0 & 0 & 0 & 0 & 0 & 0 & 0 & 0 & 0 & 0 & 9.28 & 118.18
\end{array}
\right)
\]

and
\[
\Delta D_i =
\begin{cases}
	\left( \frac{D_M}{r_d} \right)^{\mathrm{th}}(z_i) - \left( \frac{D_M}{r_d} \right)^{\mathrm{obs}}(z_i), & \text{for } D_M \text{ measurements}, \\[8pt]
	\left( \frac{D_H}{r_d} \right)^{\mathrm{th}}(z_i) - \left( \frac{D_H}{r_d} \right)^{\mathrm{obs}}(z_i), & \text{for } D_H \text{ measurements}, \\[8pt]
	
\end{cases}
\]

 \textbf{Observational Hubble Data}: The Observational Hubble data (OHD)  method relies on estimating the differential ages of the oldest passively evolving galaxies at closely separated lower redshifts. This approach provides  a model-independent determination of the expansion rate of the Universe by relating the Hubble parameter to the redshift and cosmic time via the relation
$H(z) = -\frac{1}{1+z} \frac{dz}{dt},$ as originally proposed in foundational studies~\cite{ref49}. In this work, we use 33 non-correlated measurements of the Hubble parameter $H(z)$, covering the redshift interval $0.07 \leq z \leq 1.965$ \cite{ref50,ref51,ref52,ref53,ref54,ref55,ref56,ref57}, obtained through the various Hubble surveys.\\

We define the chi-squared function for the observational Hubble data, denoted by $\chi^2_{\rm OHD}$, as follows:

$$\chi^2_{\text{OHD}} = \sum_{i=1}^{33} \frac{\left[ d^{obs}(z_i) - d^{th}(z_i) \right]^2}{\sigma^2_{d^{obs}(z_i)}},$$

Where $ d^{obs}(z_i)$ and $ d^{th}(z_i)$ denotes the observed and model-predicted values of the Hubble parameter at redshift $z_i$, respectively, and $\sigma_{{d^{obs}(z_i)}}$ represents the associated observational uncertainty. \\

\textbf{Big Bang Nucleosynthesis (BBN)}: An independent and powerful constraint on the baryon density is provided by Big Bang Nucleosynthesis (BBN), which remains one of the most sensitive probes of early-universe physics. Beyond testing standard cosmology, BBN also reveals its possible limitations. Incorporating the latest nuclear physics results from LUNA \cite{ref58}, we will obtain a precise determination of the baryon density, $\omega_b \equiv \Omega_b h^2$ in the present study.\\

 \textbf{Pantheon Plus}: The primary evidence for cosmic accelerated expansion came from Type Ia supernovae (SNIa) observations. Among the most reliable and successful probes for investigating the cosmic expansion and nature of dark energy to date are SNIa. Over the years, several high-quality supernova datasets have been compiled, with the most recent advancement being the updated Pantheon Plus sample. This compilation includes 1701 light curves corresponding to 1550 distinct SNe Ia events, covering a wide redshift range of \( z \in [0.001, 2.26] \) \cite{ref59}. Each SNIa measurement is characterized by its distance modulus, $\mu = m - M$, where $m$ denotes the observed apparent magnitude and $M$ represents the absolute magnitude of the supernova. The goodness-of-fit is then evaluated using a chi-square statistic built from the residuals between observed and theoretical values. The chi-square function is defined as: 

 \begin{equation}
	\chi^2_{\text{PP}} =  \Delta\mu \left(C_{\text{pp}}^{-1}\right) \Delta {\mu}^T.
\end{equation}

where $C_{\text{pp}}^{-1}$ denotes the inverse of the covariance matrix for PP datasets measurements, and $\Delta\mu = \mu_{th}- \mu_{obs}$, here $\mu_{th}$ and $\mu_{obs}$ represent the theoretical and observed distance moduli, respectively, $u_{th}$ is defined as, 

\begin{equation}\label{eq16}
\mu_{th} = 5 \log \left( \frac{H_0^{-1}}{1\,\text{Mpc}} \right)D_L(z) + 25,
\end{equation}

Where, the luminous distance $D_L(z)$ with non-flat cosmic curvature is defined as  

\begin{equation}\label{eq17}
D_L = (1 + z) \int_0^z \frac{H_0}{H(z')} \, dz'.
\end{equation}

Here, $H_0$ is the present value of the Hubble constant.\\ 

We consider the dynamical dark energy framework ($w$CDM) with the baseline parameter set $\mathcal{P}_{w\text{CDM}} = \{\omega_b, \omega_{cdm}, \Omega_k,  H_0, w_{deo}\}$. In our statistical analysis, we impose flat priors on all parameters: $\omega_b \in [0.018, 0.024]$, $\omega_{\rm cdm} \in [0.10, 0.14]$, $H_0 \in [60, 80]$, $\Omega_{\rm k} \in [-1, 1]$, $\Omega_{\rm m} \in [0.20, 0.35]$ and $w_{\rm de0} \in [-3, 1]$. The parameter constraints are derived through Markov Chain Monte Carlo (MCMC) sampling with a customized implementation of the \texttt{CLASS}+ \texttt{MontePython} framework \cite{ref60,ref61,ref61a}. Convergence of the MCMC chains is verified using the Gelman–Rubin criterion ($R - 1 < 0.01$)\cite{ref62}, and the posterior distributions are analyzed with the \texttt{GetDist} Python package \cite{ref63,ref63a}.

\section{Results and discussion}

In this study, we investigate the $w$CDM model with curvature using a variety of dataset combinations.  In specifically, we examine both the DR1 and DR2 dataset combinations to determine which one gives the strongest limitations on spatial curvature.  Using the DR1+BBN and DR2+BBN datasets within the $w$CDM + $\Omega_k$ framework, we find present-day curvature values of $0.094 \pm 0.080$ and $0.003 \pm 0.048$.  This suggests non-zero curvature in both combinations; however, the DR1+BBN dataset combination provides a more robust signal of curvature than the DR2+BBN combination. When the OHD data are added to both combinations, we find $\Omega_k = 0.075^{+0.070}_{-0.054}$ from DR1+BBN+OHD and $\Omega_k = 0.002 \pm 0.045$ from DR2+BBN+OHD. These findings clearly show that the DR1 + BBN + OHD combination favors a open universe, whereas the DR2 + BBN + OHD combination predicts a nearly flat but marginally open universe. The combination provides a more reliable indication of curvature than the DR2+BBN combination. From the DR1+BBN+PP and DR2+BBN+PP datasets, our results are nearly consistent with those from the first combinations. Both of these dataset combinations also suggest a non-zero cosmic curvature, favoring a open universe. As seen in Table \ref{tab1}, the cosmic curvature constraints across different dataset combinations reveal non-zero distinct values: whenever OHD data are included with the DR1+BBN+PP and DR2+BBN+PP combination, we find again positive curvature values, pointing instead towards a open universe.

\begin{figure}[hbt!]
	\centering
	\includegraphics[width=0.48\linewidth]{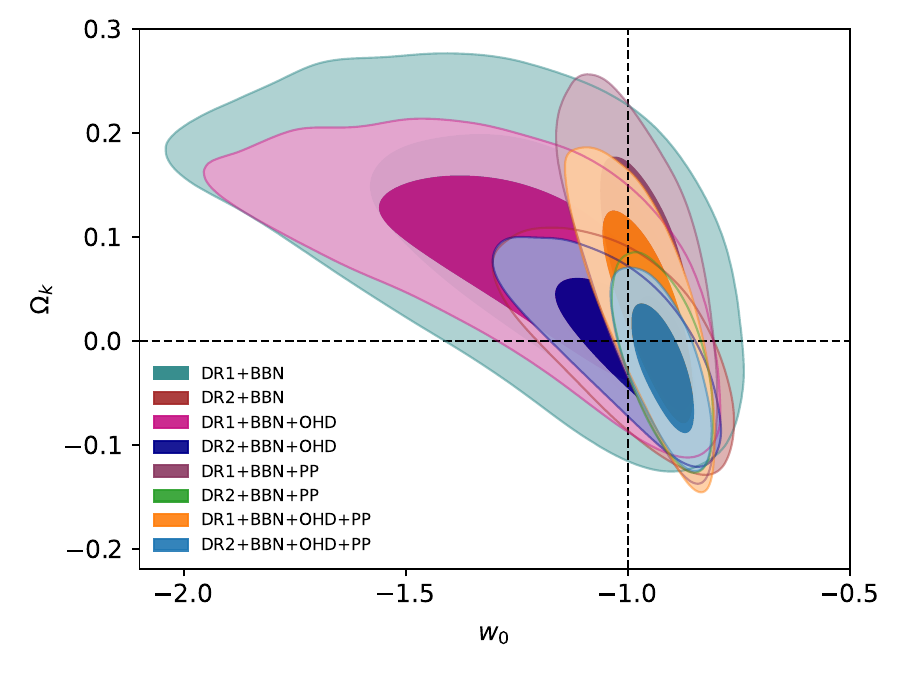}
	\includegraphics[width=0.48\linewidth]{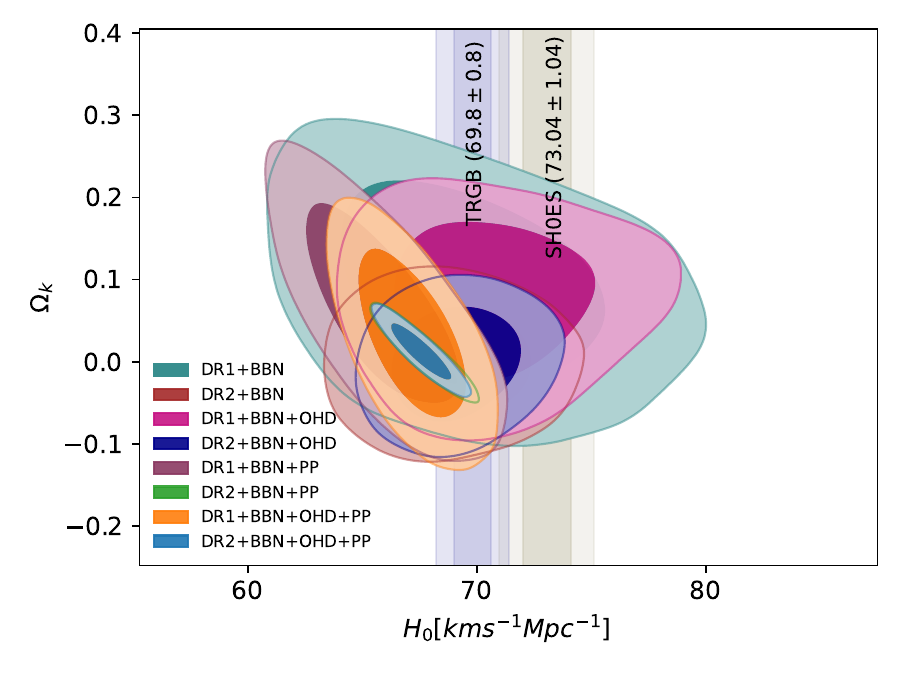}
	\caption{The left 2D contour cosmic curvature $\Omega_k$ versus EoS of DE $ w_{0}$  and the right 2D contour cosmic curvature $\Omega_k$ versus $H_0$  for $w$CDM + $\Omega_k$ with all given combination data sets. }
	\label{fig1}
\end{figure}

 From the left panel of Fig.\ref{fig1}, we clearly observe that $\Omega_k$ is negatively correlated with the dark energy equation of state parameter $w_{0}$. This implies that as the value of $\Omega_k$ decreases, the corresponding value of $w_{0}$ increases. In this figure, the horizontal black dotted line represents the flat universe ($\Omega_k = 0$), while the vertical black dotted line at $w_{0} = -1$ corresponds to the vacuum form of dark energy (cosmological constant). In the exact figure, the contours obtained from all DR1 and DR2 dataset combinations are shown to be well-constrained. We have expanded our discussion of the correlations between $w_0$ and $\Omega_k$ , which are present in the left panel of Fig.\ref{fig1}, and how they influence our results. Specifically, when we use the DR1 data combination, we obtain a noticeable deviation from spatial flatness, with the curvature parameter departing from zero. Due to the correlation shown in Fig.\ref{fig1}, it also drives the dark-energy equation-of-state parameter away from the cosmological constant,  leading to a deviation of more than a $1\sigma$ from $w_0 = -1$. However, switching from DR1 to the more precise DR2 measurements, the inferred curvature shifts toward a flat universe, and consequently, $w_0$ moves closer to the cosmological constant. We also note that when the PP dataset is included with DR2 data, the curvature parameter shows a slight deviation from flatness, indicating a mild preference for a universe that is not perfectly flat. These effects, consistent with the trends visible in the left panel of Fig.\ref{fig1}. \\

The right panel of Fig.\ref{fig1} presents the 2D contour plots of $\Omega_k$ versus $H_0$ for all combinations of DR1 and DR2 datasets, along with the $1\sigma$ level bands from SH0ES and TRGB measurements. We observe that no significant correlation between $\Omega_k$ and $H_0$ emerges in the absence of PP datasets. However, a clear negative correlation between $\Omega_k$ and $H_0$ appears when the PP dataset is included in both DR1 and DR2 combinations. Furthermore, the contours obtained from the DR2 combinations provide significantly tighter constraints compared to those from the DR1 combinations.\\

\begin{figure}[hbt!]
    \centering
    \includegraphics[width=0.7\linewidth]{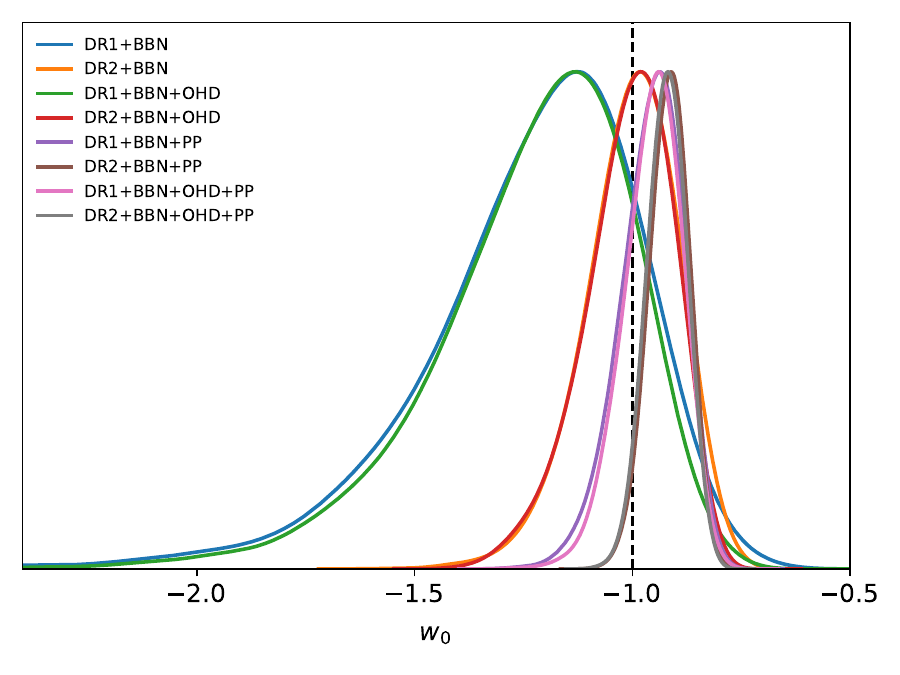}
    \caption{The posterior distribution functions (PDFs) of the EoS of DE $w_{0}$ for $w$CDM + $\Omega_k$ with all given combination data sets.}
    \label{fig2}
\end{figure}

One of the most striking puzzles in modern cosmology is the Hubble tension, arising from the disagreement between measurements of the Hubble constant, $H_0$, obtained through different methods. This tension reflects a major inconsistency between the local measurement from the SH0ES collaboration, $H_0 = 73.04 \pm 1.04~\mathrm{km s^{-1}Mpc^{-1}}$ \cite{ref64}, based on Cepheid-calibrated Type Ia supernovae, and the early-Universe estimate from the Planck collaboration, $H_0 = 67.36 \pm 0.54~\mathrm{kms^{-1}Mpc^{-1}}$ \cite{ref3}, derived from CMB observations under the $\Lambda$CDM model. The resulting discrepancy of about $5.0\sigma$ represents one of the most persistent challenges in contemporary cosmology.  Recent low-redshift observations have indicated a comparatively higher value of $H_0$ \cite{ref65,ref66} , creating a notable tension with the estimate derived from Planck-CMB data. In this section, we present the constraints on the Hubble constant for the $w$CDM +$\Omega_k$ model from different combinations of observational datasets, which are summarized in Table \ref{tab1}. From DR1+BBN datasets, we obtain a constraint of $H_0 = 69.70^{+3.60}_{-4.20}~\textcolor{blue}{\mathrm{kms^{-1}Mpc^{-1}}}$, while DR2+BBN yields $H_0 = 68.80 \pm 2.3~\textcolor{blue}{\mathrm{kms^{-1}Mpc^{-1}}}$, both of which are consistent with independent TRGB measurements. Both of these data combinations give a credible estimate of $H_0$ with bigger 1$\sigma$ uncertainties. When OHD data combined with DR1/DR2 and BBN, the 1$\sigma$ uncertainty in $H_0$ decreases slightly and also find the best-fit value of $H_0 =  70.60^{+2.50}_{-3.20}~\textcolor{blue}{\mathrm{kms^{-1}Mpc^{-1}}}$ and $H_0 = 69.10 \pm 1.7~\textcolor{blue}{\mathrm{kms^{-1}Mpc^{-1}}}$ from DR1+BBN+OHD and  DR2+BBN+OHD combination datasets, respectively. Further adding the PP dataset leads to a more stringent constraints on $H_0$, giving in $H_0 = 67.10 \pm 1.5~\textcolor{blue}{\mathrm{kms^{-1}Mpc^{-1}}}$ for DR1+BBN+OHD+PP and $H_0 = 67.60 \pm 1.2~\textcolor{blue}{\mathrm{kms^{-1}Mpc^{-1}}}$ for DR2+BBN+OHD+PP. These values are fully consistent with Planck measurements, highlighting the strong influence of the PP dataset in constraining the Hubble constant. Overall, the trend shows that as more datasets are combined, the uncertainties in $H_0$ shrink, the estimates become more precise, and the results gradually reconcile local and CMB-based measurements, with PP data playing a decisive role in this convergence.\\

\begin{table*}[hbt!]
     \caption{ The marginalized constraints, presented as mean values with 68$\%$ confidence levels (CL), on both the free and select derived parameters of the $w$CDM + $\Omega_k$ model for various DR1 and DR2 combinations.}
     \label{tab1}
     \scalebox{0.85}{
 \begin{tabular}{lcccc}
  	\hline
    \toprule
  \textbf{Model} & \multicolumn{4}{c}{\textbf{$w$CDM+$\Omega_k$}} \\  
   \\ \hline
      \textbf{Dataset} & \textbf{DR1+BBN}\,&\textbf{DR1+BBN+OHD}\,&\textbf{DR1+BBN+PP}\,&\textbf{DR1+BBN+OHD+PP}\vspace{0.1cm}\\
&\textcolor{teal}{\textbf{DR2+BBN}}\, & \textcolor{teal}{\textbf{DR2+BBN+OHD}}\, & \textcolor{teal}{\textbf{DR2+BBN+PP}}\, & \textcolor{teal}{\textbf{DR2+BBN+OHD+PP}} 
          \\ \hline

\vspace{0.1cm}
{\boldmath$10^{2}\omega_{b}$}&{$2.271\pm 0.038$ }&$2.274 \pm 0.037 $&$2.271 \pm 0.037 $ &$2.274\pm 0.038$\\
 \vspace{0.3cm}
&\textcolor{teal}{$2.271\pm 0.037$} &\textcolor{teal}{$2.272\pm 0.037$} &\textcolor{teal}{$2.271\pm 0.037$} &\textcolor{teal}{$2.273\pm 0.037$}\\

\vspace{0.1cm}
{\boldmath$\omega{}_{\rm cdm }$}&$0.1050^{+0.0160}_{-0.0200}   $ &$0.1120\pm 0.0120$ &$0.1010^{+0.014}_{-0.017}         $& $0.1100\pm 0.011$\\
 \vspace{0.3cm}
&\textcolor{teal}{$0.1160^{+0.01100}_{-0.01200}$} &\textcolor{teal}{$0.1178\pm 0.0087$} & \textcolor{teal}{$0.1136\pm 0.011$} &\textcolor{teal}{$0.1162\pm 0.0084$} \\

\vspace{0.1cm}
{\boldmath$H_0$ {[$\text{km} \text{s}^{-1} \text{Mpc}^{-1}$]}}&$69.70^{+3.60}_{-4.20}               $&$ 70.60^{+2.50}_{-3.20} $& $66.01\pm 2.2               $&$67.10\pm 1.5$ \\
 \vspace{0.3cm}
& \textcolor{teal}{$68.80\pm 2.3$}&\textcolor{teal}{$69.10\pm 1.7$}  & \textcolor{teal}{$67.30\pm 1.5$} &\textcolor{teal}{$67.60\pm 1.2$}\\

\vspace{0.1cm}
{\boldmath$w_{de0}$} &$-1.23^{+0.29}_{-0.16}             $ &$-1.24^{+0.25}_{-0.21}$ &$-0.957^{+0.084}_{-0.060}           $ & $-0.955^{+0.077}_{-0.057}$\\
 \vspace{0.3cm}
 &\textcolor{teal}{$-1.00^{+0.12}_{-0.09}$}& \textcolor{teal}{$-1.01^{+0.12}_{-0.08}$}& \textcolor{teal}{$-0.916^{+0.049}_{-0.044} $} &\textcolor{teal}{ $-0.922\pm 0.047$}  \\

 \vspace{0.1cm}
{\boldmath$M_B{\rm[mag]}$}&-  &- &$-19.483\pm 0.069$ & $-19.442\pm 0.047$\\
\vspace{0.3cm}

&- &- & \textcolor{teal}{$-19.160\pm 0.047$} &\textcolor{teal}{$-19.426\pm 0.036$}  \\

\vspace{0.1cm}
{\boldmath$\Omega{}_{m }  $}&$0.263^{+0.029}_{-0.024}$ &$0.270^{+0.025}_{-0.021}    $&$0.284\pm 0.019$& $0.295\pm 0.014  $ \\
 \vspace{0.3cm}
&\textcolor{teal}{$0.293^{+0.016}_{-0.014}   $}& \textcolor{teal}{$0.295\pm 0.014$} & \textcolor{teal}{$0.301\pm 0.012            $} &\textcolor{teal}{$0.304\pm 0.010$} \\

\vspace{0.1cm}
{\boldmath$\Omega{}_{K }  $}&$0.094\pm 0.080            $ & $0.075^{+0.070}_{-0.054}$  &$0.063\pm 0.080            $& $0.030\pm 0.068$ \\
  \vspace{0.3cm} 
&\textcolor{teal}{$0.003\pm0.048   $}& \textcolor{teal}{$0.002\pm 0.045$} & \textcolor{teal}{$0.010^{+0.012}_{-0.028}             $} &\textcolor{teal}{$0.012^{+0.015}_{-0.025}$} \\

\vspace{0.1cm}
{\boldmath$t{}_{0 }  $}&$13.97\pm 0.52            $ &{$13.75\pm 0.30$}  &$14.01\pm 0.45            $& $13.86\pm 0.29$ \\
 \vspace{0.3cm}  
&\textcolor{teal}{$13.78\pm 0.32     $}& \textcolor{teal}{$13.73\pm 0.24$} & \textcolor{teal}{$13.89\pm 0.29            $} &\textcolor{teal}{$13.81\pm 0.23$} \\

\vspace{0.1cm}
{\boldmath$z{}_{d }  $}&$1059.50\pm 1.7            $ &{$1060.00\pm 1.2$}  &$1059.10\pm 1.5            $& $1060.0\pm 1.2$ \\
  \vspace{0.3cm} 
&\textcolor{teal}{$1060.30\pm 1.2  $}& \textcolor{teal}{$1060.50\pm 1.0 $} & \textcolor{teal}{$1060.10\pm 1.2            $} &\textcolor{teal}{$1060.40\pm 1.0$} \\

\vspace{0.1cm}
{\boldmath$rs{}_{d }  $}&$151.20\pm 5.4            $ &{$149.10\pm 3.2$}  &$152.40\pm 4.6            $& $149.4\pm 3.1 $ \\
  \vspace{0.3cm} 
&\textcolor{teal}{$147.80\pm 3.1    $}& \textcolor{teal}{$147\pm 2.3$} & \textcolor{teal}{$148.5\pm 2.8            $} &\textcolor{teal}{$147.8\pm 2.3 $} \\

\hline

\vspace{0.3cm}
{\boldmath$\rm{ln} \mathcal{Z}$}&$ -6.41$&$-13.89$&$-711.69 $ &$-719.40$ \\
  
& \textcolor{teal}{$-6.78$}& \textcolor{teal}{$-14.05$} &\textcolor{teal}{$-712.04$}&\textcolor{teal}{$-719.57$}\\

\hline
\hline
\end{tabular}
}
\end{table*}

Table \ref{tab1} shows that for the DR1+BBN and DR1+BBN+OHD combinations, the best-fit value of the equation of state (EoS) of dark energy favors the phantom region (the range of EoS of DE $-1 > w_{0} $). On the other hand, for the DR2+BBN and DR2+BBN+OHD combinations, the EoS of DE is consistent with a cosmological constant ($w_0 = -1$). When both DR1 and DR2 combinations include the PP dataset, the EoS moves towards the quintessence region ($-1 < w_{0} < -1/3$). We quantify the EoS of DE obtained from $w$CDM + $\Omega_k$  model with the EoS of vacuum energy, for the DR1+BBN and DR1+BBN+OHD combinations, EoS of DE deviation from the cosmological constant by $1\sigma$. With the inclusion of the PP dataset in both DR1 and DR2 combinations, the deviation is slightly reduced to approximately $0.5\sigma$ from DR1+BBN+PP and DR1+BBN+PP+OHD, while approximately $1.8\sigma$ deviation from cosmological constant ($w_{0} = -1$) with DR2+BBN+PP and DR2+BBN+PP+OHD data sets. These results indicate that the combination of PP datasets shows a mild preference for quintessence, whereas the absence of PP data systematically pushes the EoS into the phantom regime, highlighting the influence of supernovae data in constraining the dark energy behavior.\\

Figure~\ref{fig2} displays the posterior distribution functions (PDFs) of the dark energy equation of state parameter, $w_{0}$, from all the given multiple cosmological data combinations. This figure reveals several probability distributions that correlate to various redshifts ($z$) and observational datasets. The most likely value of $w_{0}$ deduced from the corresponding dataset is shown by the peak of each distribution. The degree of consistency or tension between various metrics is depicted by the overlapping and non-overlapping zones among the distributions. Fig.~\ref{fig2}, the cosmological constant, represented by the black dotted line, is $w_{0} = -1$. Notably, the peaks of the one-dimensional distributions deviate from $w_{0} = -1$ for all given combination datasets, suggesting that our model tends to disfavour the cosmological constant scenario.\\

\begin{figure}[hbt!]
    \centering
    \includegraphics[width=0.48\linewidth]{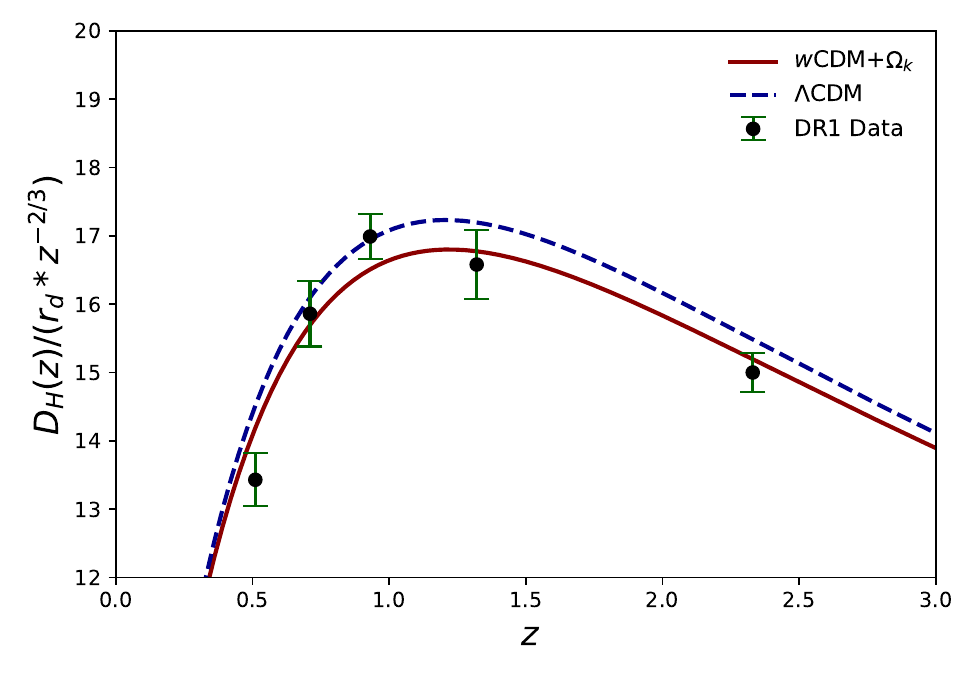}
    \includegraphics[width=0.48\linewidth]{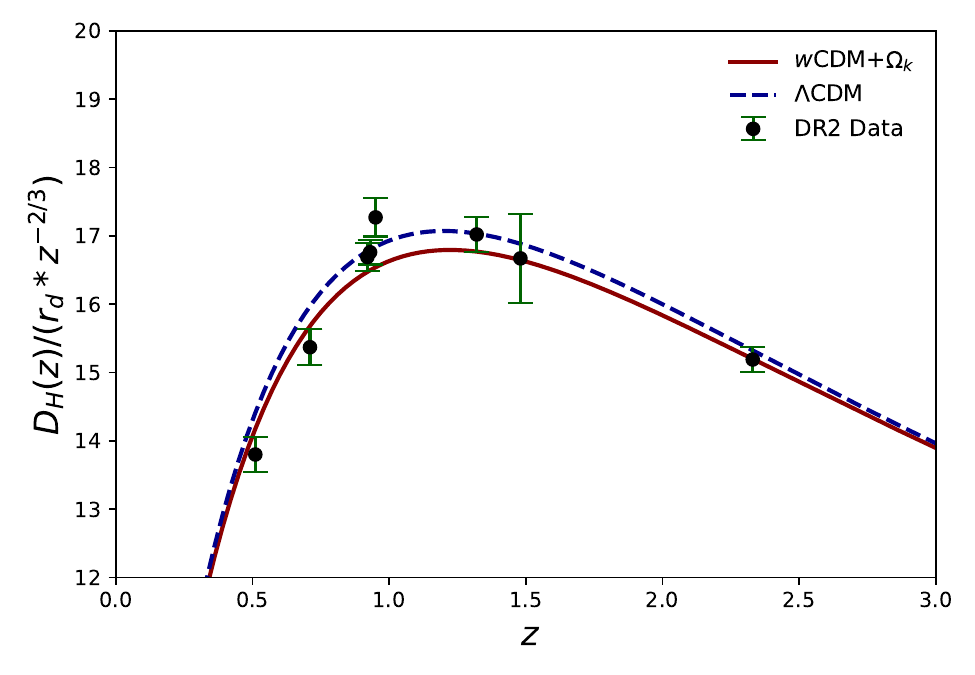}

    \caption{The 2D plot Hubble distance $D_H / (r_d *z^{-2/3})$ versus redshift z for $w$CDM + $\Omega_k$ and $\Lambda$CDM models with DR1+BBN+PP+CC and DR2+BBN+PP+CC data sets.}
    \label{fig3}
\end{figure}

In this paper, we aim to examine which of the DESI DR1 and DR2 datasets provides a better fit within our proposed model. We compare the theoretical distance predictions of the standard model with those of our model using both DR1 and DR2 data, including their respective error bars. Since the DR1 and DR2 datasets alone provide relatively weak constraints, we also use combined datasets involving CC, BBN, and PP data to strengthen the comparison of theoretical distances. Specifically, we focus on three key physical distance measures—$D_H$, $D_V$, and $D_M$—as illustrated in Fig. \ref{fig3}, \ref{fig4}, \ref{fig5}. In all figure, the black curve represents the predictions from the $\Lambda$CDM model, while the dark red curve corresponds to the $w$CDM + $\Omega_k$ model from observational combination DR1+BBN+PP+CC and DR2+BBN+PP+CC data sets, shown alongside the DR1 and DR2 data points with green error bars.\\

\begin{figure}
    \centering
   \includegraphics[width=0.48\linewidth]{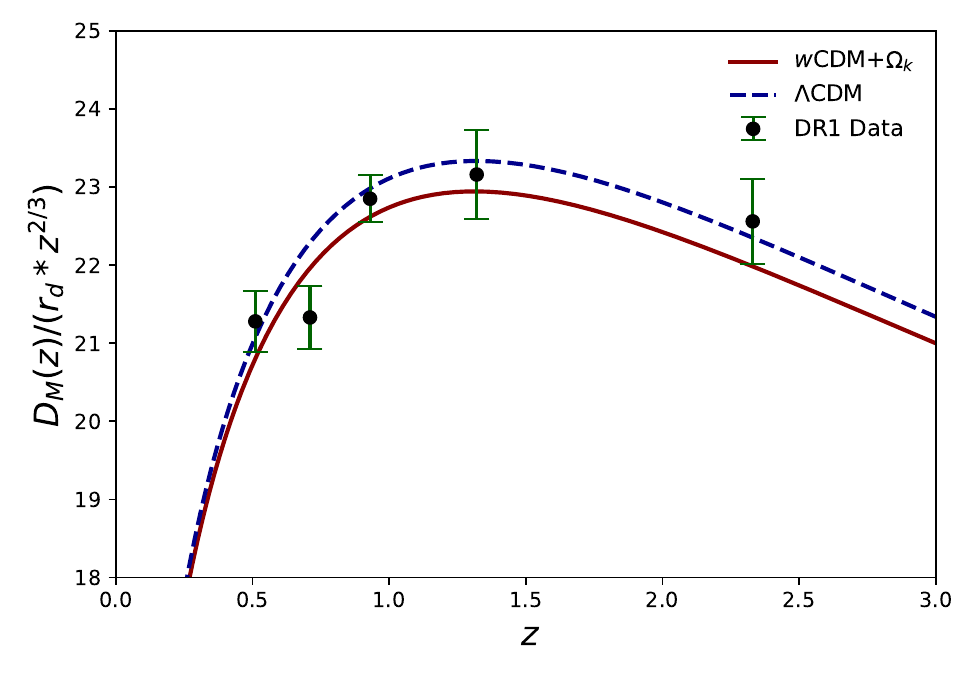}
 \includegraphics[width=0.48\linewidth]{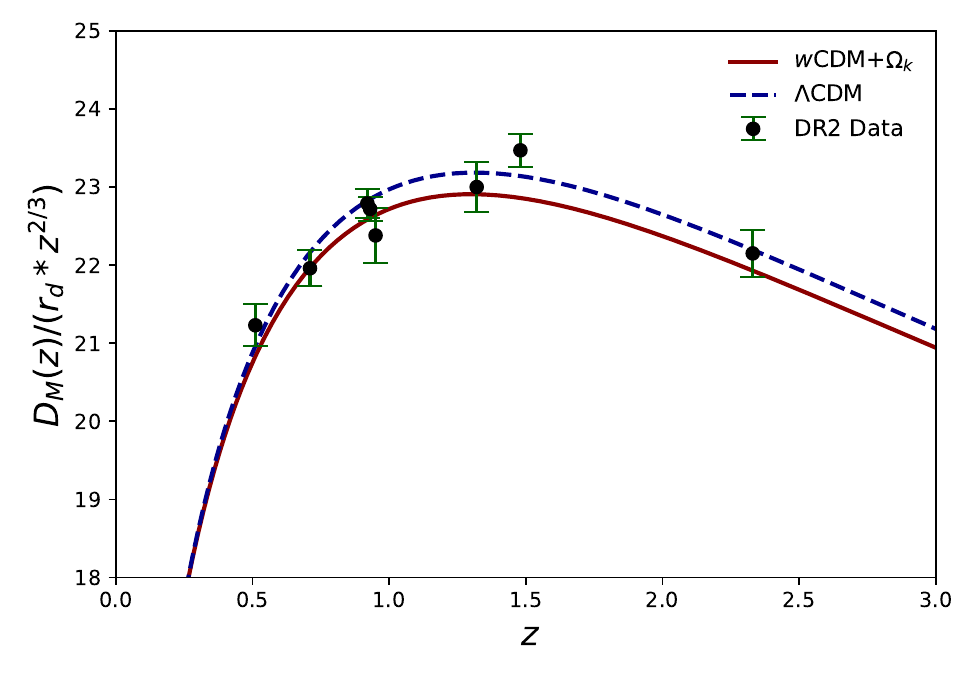}

    \caption{The 2D plot comoving angular diameter distance $D_M / (r_d *z^{2/3})$ versus redshift z for $w$CDM + $\Omega_k$ and $\Lambda$CDM models with DR1+BBN+PP+CC and DR2+BBN+PP+CC data sets.}
    \label{fig4}
\end{figure}

From this left Fig. \ref{fig3}, we observe that the curve of the standard model does not overlap with that of the dynamical model, indicating a deviation in the $w$CDM + $\Omega_k$ scenario. We first discuss the Hubble horizon distance, $D_H$, which is normalized in the Fig. \ref{fig3} as $D_H / (r_d *z^{-2/3})$. At $z = 1.32$ and $z = 2.33$, the theoretical distance predictions from  $w$CDM + $\Omega_k$  are in good agreement with the DR1 observational data points rather than the standard model. However, for the remaining low-redshift DR1 data points, both models show good consistency with the observations. When examining the DR2 data, we find that at $z = 1.32$ and $z = 2.33$, the distance predictions of both models are consistent with the DR2 measurements, as shown in the right Fig. \ref{fig3}. Meanwhile, at $z = 1.48$, $w$CDM + $\Omega_k$ model provides a better theoretical fit to the observed physical distance compared to the standard model.\\

\begin{figure}[hbt!]
    \centering
   \includegraphics[width=0.48\linewidth]{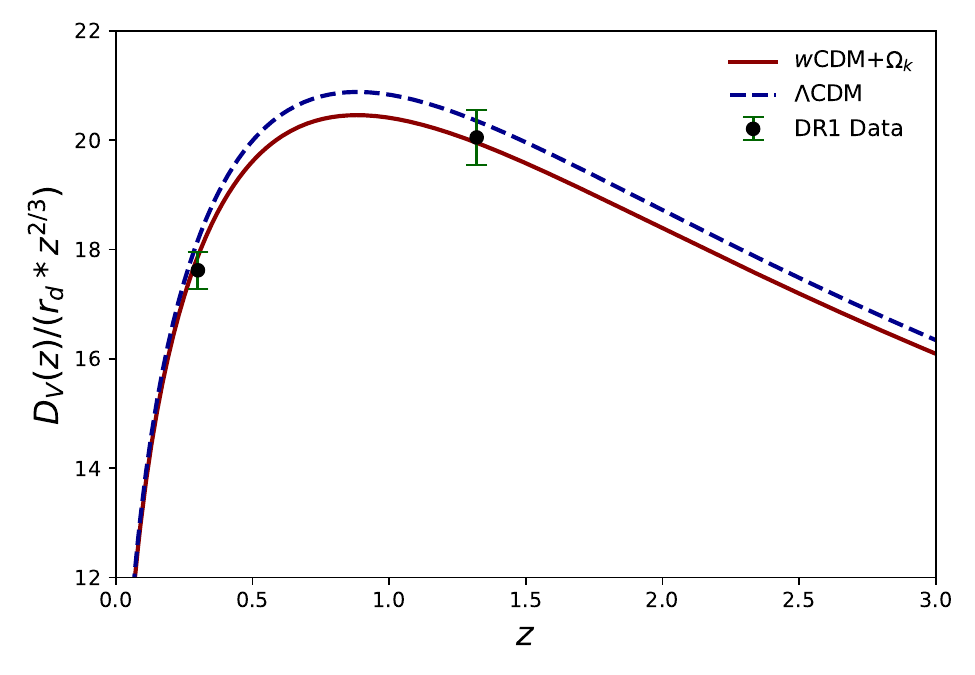}
 \includegraphics[width=0.48\linewidth]{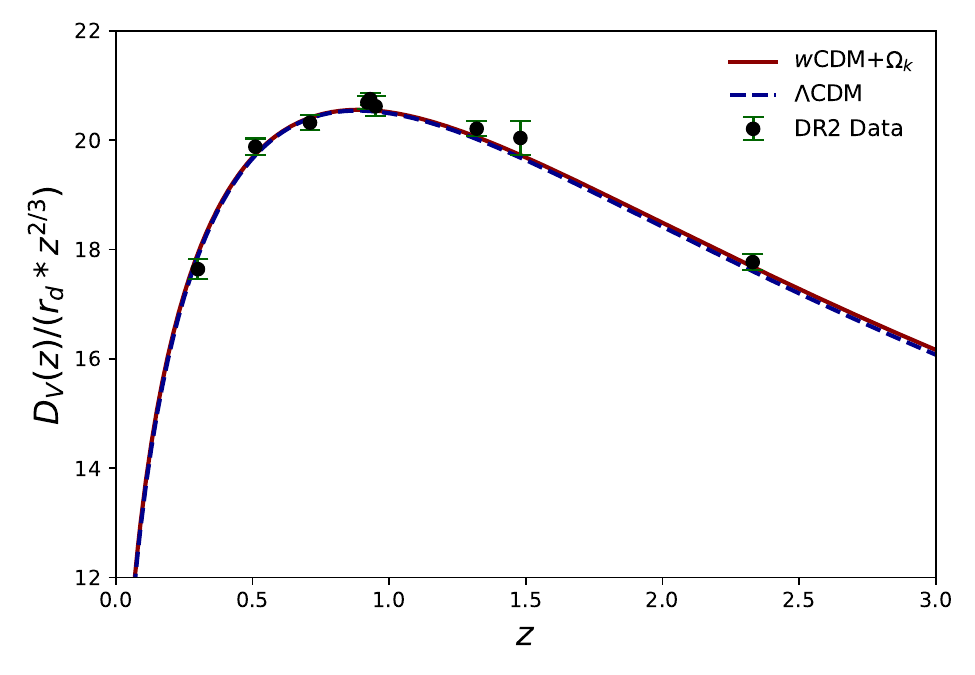}

    \caption{The 2D plot volume-averaged distance $D_V / (r_d *z^{2/3})$ versus redshift z for $w$CDM + $\Omega_k$ and $\Lambda$CDM  models with DR1+BBN+PP+CC and DR2+BBN+PP+CC data sets. }
    \label{fig5}
\end{figure}

Next, we examine the comoving angular diameter distance, $D_M$, which is normalized in the left and right Fig. \ref{fig4} as $D_M /(r_d *z^{2/3})$. At $z = 1.32$, the $w$CDM + $\Omega_k$ model shows good agreement with the physical distance inferred from  DR2 observational data points compared to the $\Lambda$CDM model. Further analysis of this figure, we notice that at $z = 0.71$, a noticeable tension remains between the theoretical predictions of both cosmological models and the DR1 observational data. However, this tension is alleviated when the DR2 data points are considered in our model, whereas the standard model continues to exhibit a slight discrepancy at the same redshift. Overall, we conclude that the distance predicted by our model is strongly consistent with the DR2 data points, while it shows relatively weaker constraints with the DR1 dataset.\\

Finally, we discuss the volume-averaged distance, $D_V$, which is normalized in the left and right Fig. \ref{fig5} as $D_V / (r_d *z^{2/3})$. Both the standard model and our $w$CDM + $\Omega_k$ model show good consistency with the volume-averaged distances inferred from the DR1 and DR2 observational data points. However, a slight tension appears at $z = 1.48$ between our model and the DR2 data point, indicating a minor deviation in the predicted distance at that redshift.

\section{Conclusions}
In this work, we have performed a comprehensive analysis of the $w$CDM$+\Omega_k$ cosmological model using the most recent spectroscopic data from the Dark Energy Spectroscopic Instrument (DESI) Data Releases 1 and 2, combined with other key cosmological probes—BBN, OHD, and Pantheon Plus (PP). Our objective is to examine the impact of spatial curvature on the constraints of cosmological parameters and to assess which DESI dataset (DR1 or DR2) provides tighter and more consistent results within the framework of dynamical dark energy.\\


Our findings show that both DESI DR1 and DR2 datasets have a small bias for a open universe, although the evidence is still below the $2\sigma$ level.  The curvature parameter $\Omega_k$ generated from the DR1+BBN and DR1+BBN+OHD combinations indicates a significantly larger divergence from flatness ($\sim1.2\sigma$) than the DR2-based combinations, which yield nearly flat but marginally open geometries. Combining DR2 data with BBN and OHD reduces uncertainty in $\Omega_k$, demonstrating the DR2 release's improved precision and resilience.  The dark energy equation-of-state parameter, $w_0$, shows a slight divergence from the cosmological constant ($w=-1$).  The DR1+BBN+OHD+PP combination produces a deviation at the $\sim0.5\sigma$ level, but the DR2+BBN+OHD+PP combination boosts this to $\sim1.8\sigma$, somewhat favoring the quintessence regime ($w>-1$). This tendency shows that the dynamical character of dark energy is consistent with present evidence but cannot be definitively differentiated from a cosmological constant. \\
%
Furthermore, the best-fit estimates of the Hubble constant ($H_0$) derived from the combined datasets are compatible with Planck 2018 results under the $\Lambda$CDM framework, spanning within the range $H_0 \approx 68\text{-}70~\mathrm{km\,s^{-1}\,Mpc^{-1}}$.  The inclusion of the OHD and PP datasets significantly reduces uncertainties in $H_0$, demonstrating the complementary role of low-redshift probes in breaking parameter degeneracy.   When examined using DR2 data, the derived key distance measures $D_H$, $D_V$, and $D_M$ show stronger internal consistency, demonstrating the increased statistical reliability of the second DESI data release. Our study shows that the DESI DR2 dataset provides improved constraints and better agreement across multiple cosmological probes within the $w$CDM$+\Omega_k$ framework. Although the evidence for non-zero curvature and dynamical dark energy remains marginal, the consistency across independent datasets strengthens the case for further investigation using upcoming high-precision DESI data releases. Future analyses incorporating CMB, weak-lensing, and gravitational-wave standard sirens will be crucial to confirm or rule out these mild deviations from spatial flatness and the cosmological constant, thereby advancing our understanding of the fundamental dynamics governing cosmic acceleration.

\section*{Declaration of competing interest}
	\noindent 
We wish to confirm that there are no known conflicts of interest
associated with this publication and there has been no significant financial
support for this work that could have influenced its outcome.

\section*{Data availability}
	\noindent 
We employed publicly available DESI BAO data, Pantheon Plus (PP) data and Observational Hubble Parameter (OHD) data presented in this study. The DESI BAO data are accessible from the official repository at https://data.desi.lbl.gov/doc/releases/. The OHD data are compiled from publicly available cosmic chronometer measurements in the literature, with a representative compilation accessible at: https://github.com/AhmadMehrabi Cosmic chronometer data. The Pantheon Plus (PP) compilation (distance moduli and covariance matrices), which is publicly available on the GitHub:https://github.com/brinckmann/montepythonpublic/tree/3.6/montepython/likelihoods/Pantheon Plus. No additional datasets are used in this study.

\begin{acknowledgments}
\noindent  
 The authors (A. Dixit \& A. Pradhan) are thankful to IUCAA, Pune, India for providing support and facility under Visiting Associateship program.  M. Yadav is sponsored by a senior Research Fellowship from the Council of Scientific and Industrial Research, Government of India (CSIR/UGC Ref.\ No.\ 180010603050). The authors are grateful to the Reviewer and Associate Editor for their fruitful remarks that increased the quality of the manuscript in its current version.   

\end{acknowledgments}

\begin{center}
   \textbf{Appendix I : Triangle Countor  } 
\end{center}

In this appendix, we present a triangular plot with 1D- and 2D- marginalized distributions for all considered parameters presented in Table \ref{tab1}  within the $w$CDM + $\Omega_k$ model based on all DR1 and DR2 combinations without PP and with PP (See Figures 6 and 7 ). 
\begin{figure}[hbt!]
    \centering
    \includegraphics[width=0.7\linewidth]{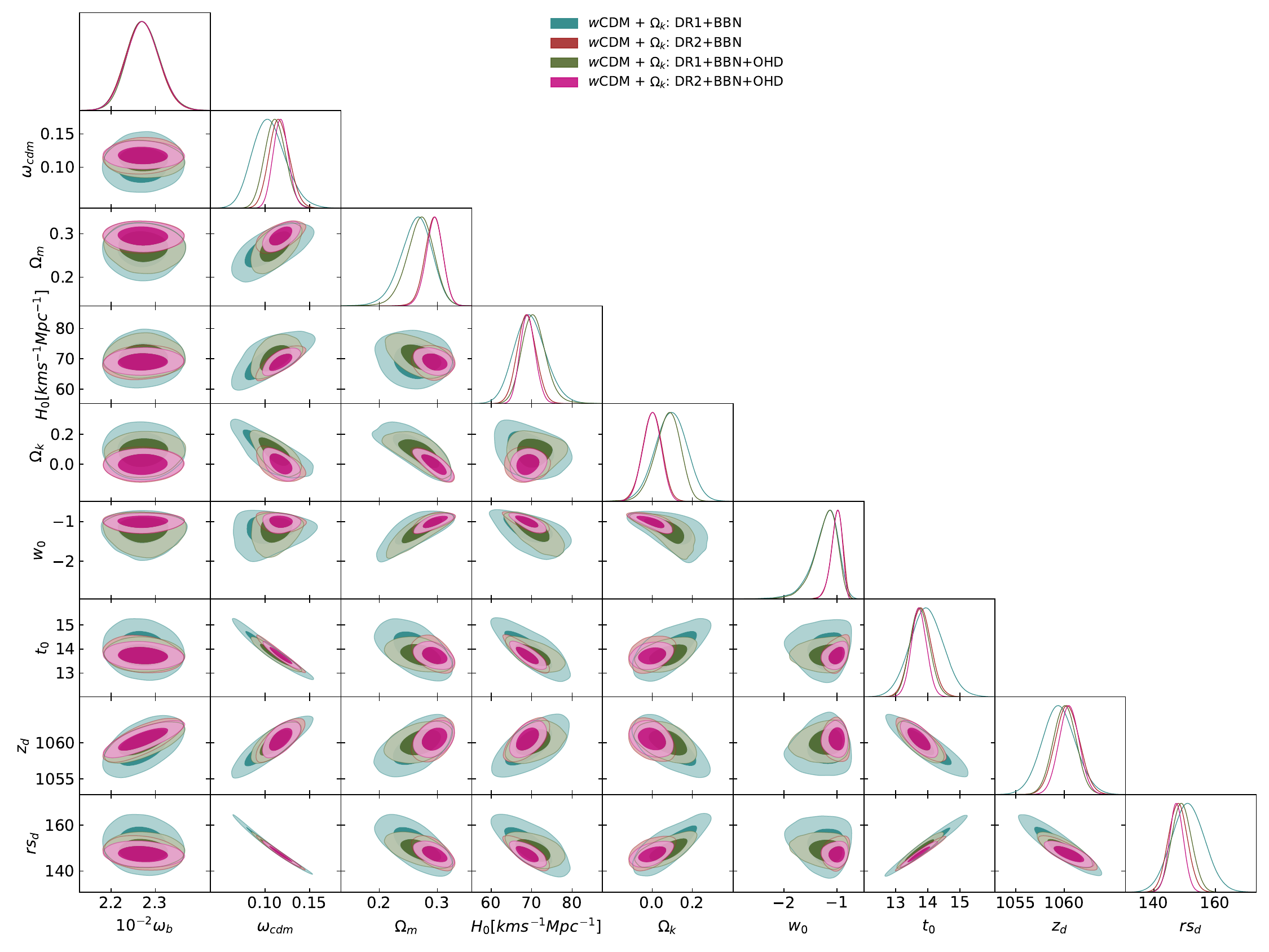}
    \caption{In this triangular plot, we present 1D- and 2D- marginalized distributions for all considered parameters presented in Table \ref{tab1}  within the $w$CDM + $\Omega_k$ model based on all DR1 and DR2 datasets combinations without PP. }
    \label{fig6}
\end{figure}

\begin{figure}[hbt!]
    \centering
    \includegraphics[width=0.7\linewidth]{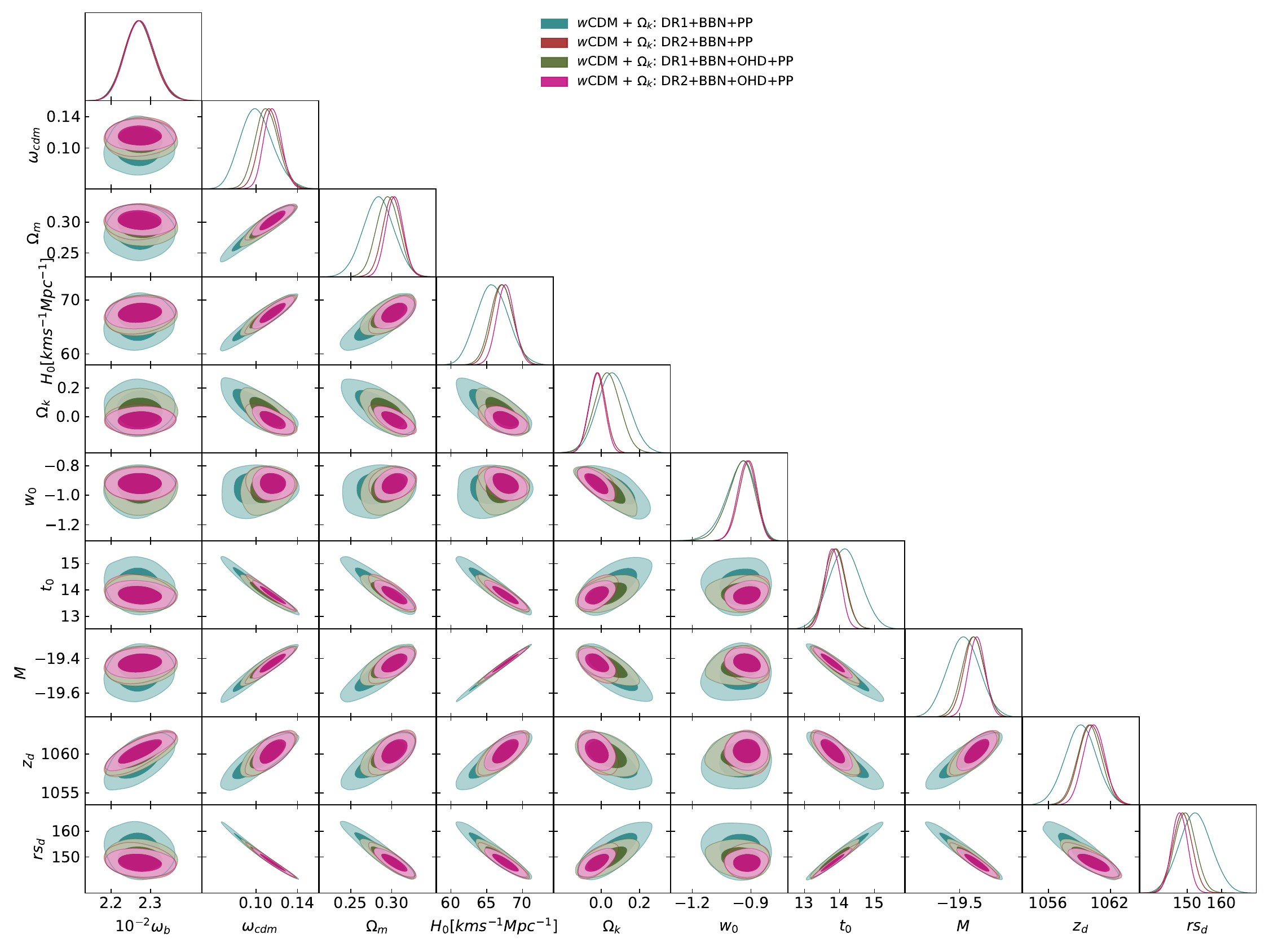}
    \caption{ In this triangular plot, we present 1D- and 2D- marginalized distributions for all considered parameters presented in Table \ref{tab1}  within the $w$CDM + $\Omega_k$ model based on all DR1 and DR2 datasets combinations with PP.}
    \label{fig7}
\end{figure}

\end{document}